# MISTAKES OF A POPULAR PROTOCOL CALCULATING PRIVATE SET INTERSECTION AND UNION CARDINALITY AND ITS CORRECTIONS

Yang Tan[1] and Bo Lv[2]

[1]Shenzhen Qianhai Xinxin Digital Technology Co.,Ltd, Shenzhen, China
`t.yang03@mail.scut.edu.cn`
[2]Huizhou University, China
`lvbo@hzu.edu.cn`

## ABSTRACT

*In 2012, De Cristofaro et al. proposed a protocol to calculate the Private Set Intersection and Union cardinality(PSI-CA and PSU-CA). This protocol's security is based on the famous DDH assumption. Since its publication, it has gained lots of popularity because of its efficiency(linear complexity in computation and communication) and concision. So far, it's still considered one of the most efficient PSI-CA protocols and the most cited(more than 170 citations) PSI-CA paper based on the Google Scholar search.*

*However, when we tried to implement this protocol, we couldn't get the correct result of the test data. Since the original paper lacks of experimental results to verify the protocol's correctness, we looked deeper into the protocol and found out it made a fundamental mistake. Needless to say, its correctness analysis and security proof are also wrong.*

*In this paper, we will point out this PSI-CA protocol's mistakes, and provide the correct version of this protocol as well as the PSI protocol developed from this protocol. We also present a new security proof and some experimental results of the corrected protocol.*

## 1. INTRODUCTION

Private Set Intersection Cardinality (PSI-CA) is an important primitive of secure two-party computation. It enables two parties, each holding a private set, to jointly compute the cardinality of their intersection without revealing any private information about their respective sets. Also, it can be extended to multi-party scenarios.

PSI-CA can be very useful in lots of scenarios. For example, in social networks(such as Facebook, Whatsapp), when two people try to determine whether they should become friends based on the number of common connections, the intuitive way is to directly exchange the information of their contacts to see if their number of common connections exceeds some threshold value. However, this method will leak private contact information. With PSI-CA, these two people can privately compute this number of common connections without revealing any private contact information.

Other useful scenarios include DNA sequence similarities comparison, anonymous authentication, etc.

In the existing PSI-CA protocols, the most efficient ones require linear computation and communication complexity. As far as we know, [1] is the first and the most famous protocol which achieves linear communication complexity. Following works tried to introduce more mechanisms such as bloom filters, homomorphic encryption[2], [3], [4], oblivious transfer [5] to improve the efficiency, but they basically stay at the same complexity level which is linear. A recent work[6] proposed a more efficient PSI-CA protocol, but this protocol has to sacrifice some accuracy for efficiency which makes it not suitable for some applications. Consequently, scholars[7], [8] are still inclined to reckon that [1] is one of the most efficient PSI-CA protocols.

However, in our implementation of this protocol, we found that the protocol can't return the correct result for the test data. Since the original paper is lack experiment results to back it up, the possibility of this protocol being wrong can't be ruled out. Thereby, we went through the details of this protocol and found a fundamental mistake it made. Needless to say, its correctness analysis and security proof are also wrong.

Since concerning the PSI-CA protocol, this paper is very influential and has the most cites(based on the Google Scholar search), we feel obliged to point out its mistakes and make corrections.

The structure of this paper is as follows. Firstly, we give a background introduction. Secondly, we describe some basic definitions. Thirdly, we describe the original PSI-CA and point out the mistakes it made and show the correct version of this protocol. We also show the correct version of the PSI protocol developed from the PSI-CA protocol. Fourthly, we give a correctness and security analysis of our new corrected protocol. Fifthly, we run a protocol simulation and implement the PSI-CA/PSU-CA, PSI protocols in a popular open-source federated learning framework: FATE[9]. Finally, we make a conclusion.

## 2. DESCRIPTION OF THE ORIGINAL PSI-CA PROTOCOL

In [1], the authors proposed a protocol to compute the Private Set Intersection Cardinality between two parties without revealing the actual private content of the set. Also, the Private Set Union Cardinality(PSU-CA) can be easily derived from the result of PSI-CA.

The authors claimed their PSI-CA protocol was secure under DDH assumption[10] in the random oracle model against semi-honest adversaries.

### 2.1. Definitions

Firstly, we introduced some basic definitions of the original PSI-CA.

**Server** A party in this paper is referred to as **S** with a private input set $s_1, ..., s_w$.

**Client** A party in this paper is referred to as **C** with a private input set $c_1, ..., c_v$. This is the party that sends a PSI-CA/PSU-CA request to the Server and it will get the final result.

**Private Set Union Cardinality(PSU-CA)** A protocol involving server, on input a set of $w$ items $S = \{s_1, ..., s_w\}$, and client, on input a set of $v$ items $C = \{c_1, ..., c_v\}$. It outputs $|U|$, where: $U = S \cup C$.

**Private Set Intersection Cardinality(PSI-CA)** A protocol involving server, on input a set of $w$ items $S = \{s_1,...,s_w\}$, and client, on input a set of $v$ items $C = \{c_1,...,c_v\}$. It outputs $|I|$, where: $I = S \cap C$.

**Private Set Intersection (PSI)** A protocol involving server, on input a set of $w$ items $S = \{s_1,...,s_w\}$, and client, on input a set of $v$ items $C = \{c_1,...,c_v\}$. It outputs $I$, where: $I = S \cap C$.

For both PSI-CA and PSU-CA protocols, the following privacy requirements should be met:
- **Server Privacy** Client learns no information beyond: (1) cardinality of set intersection/union and (2) upper bound on the size of S.
- **Client Privacy** No information is leaked about client set C, except an upper bound on its size.
- **Unlinkability** Neither party can determine if any two instances of the protocol are related, i.e., executed on the same input by client or server, unless this can be inferred from the actual protocol output.

One thing to note, for any set $C$ and $S$, the size of the union $C \cup S$ can be computed as $|C|+|S|-|C \cap S|$. Thereby, with the result of PSI-CA, one can easily get PSU-CA.

Other definitions involved in this protocol:
- Two hash functions act as random oracles, $H1:\{0,1\}^* \to Z_p^*$ and $H2:\{0,1\}^* \to \{0,1\}^k$ given the security parameter $k$. $H1$ and $H2$ are both deterministic which means if $x = y$, we have $H1(x) = H1(y)$ and $H2(x) = H2(y)$.
- Two random data permutations $\prod, \prod'$ which randomly shuffles the item order of a data set. All the calculations will happen in $G$ which is a cyclic group of order $q$ and with $g$ as its generator.

## 2.2. Original Protocol

Here in Figure 1, shows the original protocol described in [1].

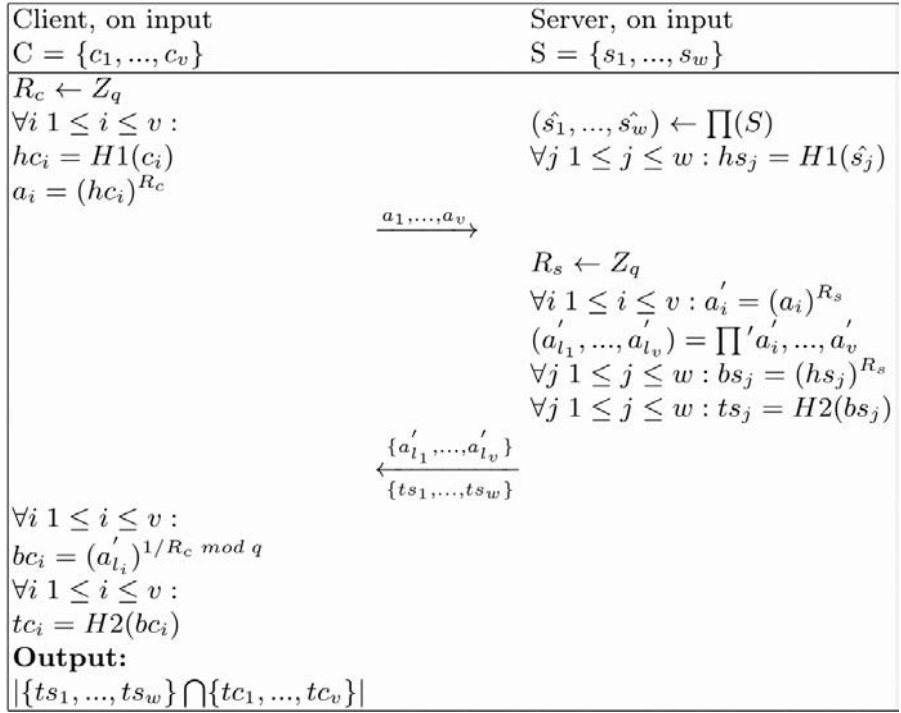

Figure 1. Original PSI-CA Protocol from [1].

The original protocol's idea comes from the following intuition.

**Intuition.** First, the client masks its set items $c_i$ with hash function $H1$ and a random exponent ($R_c$) generated on his side and sends the resulting values ($a_i$) to the server which further masks them by exponentiating them with its own random value $R_s$. The server randomly shuffles(i.e. $\prod'$) these values to prevent the client from recovering the exact intersecting item by item's order. The resulting values $a'_{l_i}$ after shuffle will be sent back to the client.

Then, the server masks its own set items $s_i$ by the following order: A new random shuffle $\prod$; hash function $H1$; exponentiating them with its own random value $R_s$; hash function $H2$. The resulting values $ts_i$ will be sent to the client for client to do the final PSI-CA calculation.

When client receives further randomly shuffled and exponentiated items $a'_{l_i}$, he will strip of the initial exponent ($R_c$) by exponentiating $R_c$'s reverse modular $q$. He then further masks the set items with the hash function $H2$ and outputs $tc_i$.

From the process description, we can see that, from the client's side, in the end, the data he gets $tc_i$, $ts_i$ are the client and server's private set items which went through the same calculations: hash function $H1$ and exponentiating with $R_s$ and hash function $H2$ and along with some random order shuffles in the process which won't affect the actual value but the order. Thereby, if $c_i = s_j$, then we have $tc_i = ts_i$ and the client can do the intersect cardinality computation with simple equality checks between these two data sets $\{tc_i\}$, $\{ts_j\}$. However, the client cannot recover the original private item under the DDH assumption since he doesn't know the random value $R_s$ and how the items are shuffled ($\prod$, $\prod'$).

To draw a conclusion, the client finally gets the correct result of PSI-CA without knowing the actual intersection.

## 2.3. Why is it wrong and its corrections

From the protocol description and the intuition behind it, the original protocol seems alright. However it made a fundamental mistake: all the computations, including hash functions and exponentiations with random values, didn't happen in the chosen cyclic group. They are just normal computations in modular $p$. Without mapping values to the chosen group, its security dependency: DDH assumption won't stand under this circumstance.

To be more specific, computations like stripping off the initial exponent $R_c$ by exponentiating with $R_c$'s reverse modular $q$ won't work.

After figuring out what's wrong with the original protocol, we correct it by mapping the intermediate results(values after being processed by hash function $H1$) to the chosen cyclic group on both the Client and the Server sides. Specifically, we do more exponentiations with generator $g$ as the base, and intermediate results as the exponents. With this correction, the client can successfully strip the initial exponent($R_c$) off now.

In Figure 2, we show the description of the full corrected protocol.

```
Client, on input                                    Server, on input
C = {c_1, ..., c_v}                                 S = {s_1, ..., s_w}
─────────────────────────────────────────────────────────────────────
R_c ← Z_q
∀i 1 ≤ i ≤ v :                                      (ŝ_1, ..., ŝ_w) ← ∏(S)
hc_i = H1(c_i)                                      ∀j 1 ≤ j ≤ w :
hgc_i = g^{hc_i}                                    hs_j = H1(ŝ_j);
a_i = (hgc_i)^{R_c}                                 hgs_j = g^{hs_j}
                        ──a_1,...,a_v──▶

                                                    R_s ← Z_q
                                                    ∀i 1 ≤ i ≤ v : a'_i = (a_i)^{R_s}
                                                    (a'_{l_1}, ..., a'_{l_v}) = ∏'a'_i, ..., a'_v
                                                    ∀j 1 ≤ j ≤ w :
                                                    bs_j = (hgs_j)^{R_s};
                                                    ∀j 1 ≤ j ≤ w : ts_j = H2(bs_j)

                        ◀──{a'_{l_1},...,a'_{l_v}}──
                           {ts_1,...,ts_w}

∀i 1 ≤ i ≤ v :
bc_i = (a'_{l_i})^{1/R_c mod q}
∀i 1 ≤ i ≤ v :
tc_i = H2(bc_i)
Output:
|{ts_1, ..., ts_w} ∩ {tc_1, ..., tc_v}|
```

Figure 2. Correct version of the PSI-CA Protocol from [1].

**Complexity.** The correct version of protocol's complexity remains the same level (linear complexity) as the original paper [1] claimed which is $O(w+v)$ computation and communication. Specifically, according to the protocol description in Figure 2, the client performs $3v$ exponentiations with $|q|$−bit exponent and $|p|$−bit modular, $2v$ hash functions and send $v$ data items to server. On the other side, the server performs $2w + v$ exponentiations with $|q|$−bit exponent and $|p|$−bit modular, $2w$ hash functions and sends $v + w$ data items to the client.

## 2.4. Corrections on the extended PSI protocols

In the original paper, based on the PSI-CA protocol, the authors extended two additional protocols.

One is called authorized PSI-CA(APSI-CA) in which client's input must be pre-authorized by an off-line mutually-trusted authority. In this protocol, all client's input $c_1,c_2,...,c_v$ will be attached with an RSA signature $\sigma_1,\sigma_2,...,\sigma_v$ released by the mutually trusted authority on its hashed value (with hash function $H1$). This time, the authors didn't make a mistake since this protocol's security no longer relies on the DDH assumption but on RSA[11]. Also, we checked the protocol's mathematical deduction. No correction is needed.

The other one is PSI protocol which was directly derived from the original PSI-CA protocol with only one additional round of communication. Ergo, the PSI protocol made the same mistakes as PSI-CA protocol. Thereby, it also needs correction. We skipped the description of the wrong PSI protocol. In Figure 3, we directly give the correct version of the extended PSI protocol. Compared to the PSI-CA protocol, we can see that PSI has an extra last step in which the client sends the intersection of $\{tc_i\}$ and $\{ts_j\}$ to the server for the server to do the final matching and output the PSI since server can deduce the original item from $ts_j$. Besides that, all the steps are the same with PSI-CA. Ergo, in Figure 3, we make the same correction as PSI-CA.

```
Client, on input                                     Server, on input
C = {c_1, ..., c_v}                                  S = {s_1, ..., s_w}
─────────────────────────────────────────────────────────────────────
R_c ← Z_q
                                                     (ŝ_1, ..., ŝ_w) ← ∏(S)
∀i 1 ≤ i ≤ v :                                       ∀j 1 ≤ j ≤ w :
hc_i = H1(c_i)                                       hs_j = H1(ŝ_j);
hgc_i = g^{hc_i}                                     hgs_j = g^{hs_j}
a_i = (hgc_i)^{R_c}
                          a_1,...,a_v
                          ──────────→

                                                     R_s ← Z_q
                                                     ∀i 1 ≤ i ≤ v : a'_i = (a_i)^{R_s}
                                                     (a'_{l_1}, ..., a'_{l_v}) = ∏'a'_i, ..., a'_v
                                                     ∀j 1 ≤ j ≤ w :
                                                     bs_j = (hgs_j)^{R_s};
                                                     ∀j 1 ≤ j ≤ w : ts_j = H2(bs_j)
                          {a'_{l_1},...,a'_{l_v}}
                          {ts_1,...,ts_w}
                          ←──────────
∀i 1 ≤ i ≤ v :
bc_i = (a'_{l_i})^{1/R_c mod q}
∀i 1 ≤ i ≤ v :
tc_i = H2(bc_i)
T = {ts_1, ..., ts_w} ∩ {tc_1, ..., tc_v}
                                             T
If PolicyIsSatisfied(w,v,|T|) :  ──────────→         ∀tc_j ∈ T : Output c_j ∈ S ∩ C
```

Figure 3. Correct version of the PSI Protocol from [1].

**Remark 1:** The authors seemed to forget to include the calculations of $ts_j$ in the PSI protocol description in [1]'s Figure 3, we also correct this part.

**Remark 2:** *PolicyIsSatisfied* are some policies for the client to decide whether to proceed with PSI protocol based on the PSI-CA result and $w, v$. For example, PSI-CA result must exceed a threshold value(e.g. $0.8v$) to proceed PSI. Also, the roles of server and client are not fixed. They can be reversed depending on the specific application scenario.

## 3. CORRECTNESS AND SECURITY ANALYSIS

In this section, we will give the correctness and security proof of our corrected protocol.

### 3.1. Correctness

For any $c_i$ held by client and $s_j$ held by server, if $c_i = s_j$, hence, $hc_i = hs_j$, we obtain:

$$
\begin{aligned}
&|\{ts_1,\ldots,ts_w\} \cap \{tc_1,\ldots,tc_v\}| \\
=& |\{bs_1,\ldots,bs_w\} \cap \{bc_1,\ldots,bc_v\}| \\
=& |\{g^{hs_1 \cdot R_s},\ldots,g^{hs_w \cdot R_s}\} \cap \{g^{hc_1 \cdot R_s},\ldots,g^{hc_v \cdot R_s}\}| \\
=& |\{g^{hs_1},\ldots,g^{hs_w}\} \cap \{g^{hc_1},\ldots,g^{hc_v}\}| \\
=& |S \cap C|
\end{aligned}
\quad (1)
$$

hence, client learns set intersection cardinality by counting the number of matching pairs ($tc_i$, $ts_j$).

### 3.2. Security Analysis

In this section, we will define the threat model and some security assumptions related to our protocol. Then we give the security proof of the corrected PSI-CA protocol. The PSU-CA and PSI protocol's security proofs are skipped since they are directly derived from the PSI-CA protocol and their security proof are basically the same as PSI-CA's.

#### 3.2.1 Threat Model

Firstly, we define the threat model and classes of adversaries of our PSI-CA, PSU-CA, PSI protocols.

**Semi-honest Security** In the Semi-honest Security threat model, the protocol participants will always follow the protocol's procedure. However, they are curious and will try to gain extra information out of the protocol.

After giving the definition of the threat model, we can now define two classes of adversaries concerning the Client privacy and Server privacy security requirements defined in the previous section.

**Client adversary** The client participant of the protocol who tries to violate the Server privacy. With the data he gets on his side, he will try to recover server's private items or intersection items.

**Server adversary** The server participant of the protocol who tries to violate the Client privacy. With the data he gets on his side, he will try to recover client's private items or intersection cardinality.

One thing to note, adversaries outside of the protocol are not considered since they can't gain more information than the protocol participants. If the more powerful participant adversary can't break the protocol, outsider adversaries can't break the protocol either.

### 3.2.2 Security Assumptions

**Discrete Logrithm Assumption** Let $G$ be a cyclic group and $g$ be its generator. The discrete logarithm problem(DLP) is called $(t, \varepsilon)$ hard relative to $G$ if for all algorithms $A$ runs in time $t$ there exists a negligible function $\varepsilon$ of security parameter $k$ such that

$$Pr[\mathcal{A}(g, g^a) = a] \leq \varepsilon \quad (2)$$

**DDH assumption** Let $G$ be a cyclic group of order $q$ and $g$ is its generator and $l$ is the bit-length of the group size. The following two distributions are computationally indistinguishable
(1) $g^x, g^y, g^{xy}$
(2) $g^x, g^y, g^z$
given $x, y, z$ are randomly and independently chosen from $Z_q$. To put it in a more formal way, DDH problem is $(t, \varepsilon)$ hard if for all algorithms $A$ runs in time t there exists a negligible function $\varepsilon$.

$$|\Pr[x, y \leftarrow \{0,1\}^l : A(g, g^x, g^y, g^{xy}) = 1] - $$
$$\Pr[x, y \leftarrow \{0,1\}^l : A(g, g^x, g^y, g^z) = 1]| \leq \varepsilon \quad (3)$$

If we consider more powerful adversaries, we have following security assumptions:

**One-More-DH assumption[12]** DH problem is hard even if the adversary is given access to a $DH_x(.)$ oracle (i.e. given random $h$ in group $G$, the oracle return $h^x$). Formally, let $(G, q, g) \leftarrow KeyGen(k)$, and $x \leftarrow Z_q$, we say One-More-DH is $(t, \varepsilon)$ hard for all algorithm A runs in time $t$ there exists a negligible function $\varepsilon$:

$$\Pr[\{(g_i, (g_i)^x)\}_{i=1,\dots,v+1} \leftarrow A^{DH_x(\cdot)}(g_1, \dots, g_{ch})] \leq \varepsilon \quad (4)$$

where $ch \geqslant v$ and $A$ can make at most $v$ queries to $DH_x(.)$.

### 3.2.3 Security Proof

a) *Client Privacy:*

According to the threat model, the Server adversary will try to violate the Client Privacy. According to the protocol description, the data he will receive from the client is $\{a_i\}$ where $a_i = (hgc_i)^{R_c}$ in which $hgc_i = g^{hc_i}$, $hc_i = H1(c_i)$.

The private items the Server adversary tries to recover from the Client can be divided into the following classes:

- **Non-intersection Data** If no server item satisfies $s_j = c_i$, because of the hardness of the DLP and $a_i$ $a_i = g^{hc_i \times R_c}$, server adversary can't find an algorithm $A$ that runs in $(t, \varepsilon)$ to recover $hc_i \times R_c$. Even if we discard the DLP assumption, and let the server adversary gets $hc_i \times R_c$, without the knowledge of random $R_c$, $hc_i \times R_c$ is indistinguishable from a random $r \leftarrow Z_q$. Thereby, under this circumstance, Client Privacy is guaranteed.

- **Intersection Data** If server adversary has item satisfies $s_j = c_i$, then the server adversary has $a_i = g^{hc_i \times R_c}$ and some $g^{hs_j} = g^{hc_i}$. We consider the following casese:
(1) If the intersection cardinality equals 0, i.e. the private intersection set is empty, the security proof is the same as Non-intersection Data case.

(2) If the intersection cardinality equals 1 and let's assume the intersecting item is $c_i$, even if we discard the DLP assumption let server adversary gets $hc_i \times R_c$ from $a_i$, without the knowledge of $R_c$, $hc_i \times R_c = hs_j \times R_c$ is indistinguishable from a random $r \leftarrow Z_q$, he can't recover the $hs_j$ to get the corresponding intersection item.

(3) If the intersection cardinality is greater than 1, let's assume it's $n$, and the intersection item is $s_{j1},...,s_{jn}$, then the server adversary has $g_1 = g^{hs_{j1}},...,g_n = g^{hs_{jn}}$ along with $a_{j1} = g_1^{R_c}, a_{j2} = g_2^{R_c},..., a_{jn} = g_n^{R_c}$. This case matches the One-More-DH assumption. According to this assumption, even if we give the server adversary access to $DH_x(.)$ oracle and let it recover $[0, n-1]$ pairs of $(g_1, a_{j1}), (g_2, a_{j2}),...,(g_n, a_{jn})$, it can't find an algorithm $A$ runs in $(t, \varepsilon)$ to recover another pair of $(g_t, a_{jt})$ where $1 \leqslant t \leqslant n$ and use $g_t$ to get the corresponding intersection item. Thereby, the client privacy is also guaranteed for this case under One-More-DH assumption.

Thereby, for both Intersection Data and Non-intersection Data, Client privacy is guaranteed.

b) *Server Privacy:*

According to the threat model, the client adversary will try to violate the Server Privacy. According to the protocol description, the data he will receive from the server is $\{ts_j\}$ and $\{a'_{l_i}\}$ where $ts_j = H2((hgs_j)^{R_s}) = H2(g^{hs_j \times Rs})$ and $a'_{l_i} = g^{hc_i \times R_c \times R_s}$.

For the simplicity of the proof calculations, the random permutations that prevent the client from recovering intersecting items based on the item's order are not concluded in the calculations. As long as these permutations are random, recovering items order has no advantage over blindly picking items from client's data set. Thereby, it's safe from order- recovering attack.

Similar to the previous case, the private items the client adversary tries to recover from the server can be divided into the following classes:
- **Non-intersection Data** If no Client item satisfies $c_i = s_j$, since $H2$ is modeled as random oracle, $ts_j$ is indistinguishable from random $r \leftarrow Z_p$. Even if the adversary somehow reverses the $H2$ and gets $(hgs_j)^{R_s} = g^{hs_j \times Rs}$, because of the hardness of the DLP, client adversary can't find an algorithm $A$ that runs in $(t, \varepsilon)$ to recover $hs_j \times R_s$. Even if we discard the DLP assumption and let server adversary gets $hs_j \times R_s$, without the knowledge of $R_s$, $hs_j \times R_s$ is indistinguishable from a random $r \leftarrow Z_q$. Thereby, under this circumstance, Server Privacy is guaranteed.

- **Intersection Data** If the client adversary has item that satisfies $c_i = s_j$, for this case, $H2$ and $R_c$ can both be stripped of by client adversary to perform the attack. The client adversary will have $bc_i = g^{hc_i \times R_s} = g^{hs_j \times R_s}$ and some $g^{hc_i} = g^{hs_j}$

  We consider the following cases:
(1) If the intersection cardinality equals 0, i.e. the private intersection set is empty, the security proof is the same as Non-intersection Data case above.

(2) If the intersection cardinality equals 1 and assume the intersecting item is $c_i = s_j$. Because of the DLP assumption, client adversary can't get $hc_i \times R_s$ or the specific $g^{hc_i} = g^{hs_j}$ without the knowledge of $R_s$. Even if we discard the DLP assumption let client adversary gets $hs_j \times R_s$, without the knowledge of $R_s$, $hs_j \times R_s = hc_i \times R_s$ is indistinguishable from a random $r \leftarrow Z_q$. Thereby, he can't recover $hc_i$ to get the corresponding intersection item.

(3) If the intersection cardinality is greater than 1, let's assume it's $n$, and the intersection item is $c_{i1} = s_{j1}, ..., c_{in} = s_{jn}$, then the client adversary has $g_1 = g^{hc_{i1}}, ..., g_n = g^{hc_{in}}$ along with $bc_{i1} = g_1^{R_s}, bc_{i2} = g_2^{R_s}, ..., bc_{in} = g_n^{R_s}$. This case also matches the One-More-DH assumption. According to this assumption, even if we give the client adversary access to $DH_x(.)$ oracle and let it recover $[0, n-1]$ pairs $(g_1, bc_{i1}), (g_2, bc_{i2}), ..., (g_n, bc_{in})$, it can't find an algorithm $A$ runs in $(t, \varepsilon)$ to recover one more pair of $(g_t, bc_{jt})$ where $1 \leqslant t \leqslant n$ and use $g_t$ to get the corresponding intersection item. Thereby, the server privacy is also guaranteed for this case under One-More-DH assumption.

Thereby, for both Intersection Data and Non-intersection Data, Server privacy is guaranteed.

## 4. PROTOCOL SIMULATION AND IMPLEMENTATION

After we give the correct descriptions of the PSI and PSI-CA protocol, in this section, we will make protocol simulations and implementations.

Firstly, we run a simple python simulation for the PSI-CA protocol. The python environment we use for simulation is Python 3.6.8 and we run this simulation on a Linux Red Hat 4.8.5-44 server with Intel Cure2 Duo T7700(2.4GHz) as CPU.

The parameter for the cyclic group comes from the [13]: **1024-bit MODP Group with 160-bit Prime Order Sub-group** which means a 1024-bit $p$ and 160-bit $q$.

SHA256 with different salts are used to construct $H1$ and $H2$ separately.

The initial parameters for client and server are illustrated in Figure 4 (Only showed a part of $p$ because of its great length).

```
1024bit-p:124325339146889384540494091085456630009856882741872806181731279018491820800119460022367403769795008250021191767583423221479
160bit-q:1399252811935680595399801714158014275474696840019
client's random value rc:9021590830740631550026620782167207842288316912
server's random value rs:4540526189697674679310201663852192221720023286777
client c_i: 3
client c_i: 4
client c_i: 5
client c_i: 2
client c_i: 6
server s_j: 3
server s_j: 4
server s_j: 5
server s_j: 7
```

Figure 4. Initial Parameters For PSI-CA.

The intermediate results for $tc_i$ and $ts_j$ and final outcome are illustrated in Figure 5.

```
ts_j 00f55c196284695cf1646580471bd3aff8cb4e570127db08878640150812b677
ts_j 18e95e7c065a7f51749174d26fb9b458edd1246b602d6c419059fbed9eab17c4
ts_j 7ab697f79c32457c8a337e2b4c66770ff1fa7dd85e614414400a4c906081a7cd
ts_j ac183e2d6c737cda6429e8b92e9f511f6418fb981c71252756afe71b67930602
tc_i: 00f55c196284695cf1646580471bd3aff8cb4e570127db08878640150812b677
tc_i: 4da9cb488691c688baa3076522ba42b62e70c3dfd12169baeef9d83c3d0fc086
tc_i: 6a406977a818d6e0382d080f306441e45946b7421d97087cc96710dbaf3c0b9a
tc_i: 7ab697f79c32457c8a337e2b4c66770ff1fa7dd85e614414400a4c906081a7cd
tc_i: ac183e2d6c737cda6429e8b92e9f511f6418fb981c71252756afe71b67930602
intersect of tc_i and ts_j 00f55c196284695cf1646580471bd3aff8cb4e570127db08878640150812b677
intersect of tc_i and ts_j 7ab697f79c32457c8a337e2b4c66770ff1fa7dd85e614414400a4c906081a7cd
intersect of tc_i and ts_j ac183e2d6c737cda6429e8b92e9f511f6418fb981c71252756afe71b67930602
PSI_CA: 3
```

Figure 5. Intermediate Results $tc_i$, $ts_j$ and Final Output for PSI-CA.

The simulation proves our new version of PSI-CA protocol is correct.

We didn't do the python simulation for PSI. Instead, we directly implement PSI-CA, PSU-CA, and PSI protocols in a very popular(3.1k stars and 900 forks) open-source multi- party federated learning framework FATE[9] in which PSI is very common operations for different parties to perform Sample-Aligned Federated Learning. Since this framework doesn't support PSI-CA, our work can be a pretty useful supplementary for FATE.

In FATE, there are three different parties: Guest, Host and Arbiter. In our implementation, the Guest party will play the role of client which will get the result of PSI-CA, PSU-CA. The Host party will play the role of server which can get the PSI result if the Guest agrees to proceed with the PSI computation. The Arbiter is not needed in our protocols. Both Guest and Host can choose to share their side's result with another party by setting parameters in a configuration file.

For the illustration, we use the data of "epsilon_5k_hetero_guest.csv", "epsilon_5k_hetero _host.csv" to run a test on the same server as we run PSI-CA simulation. We deploy FATE in the Stand-alone Host mode and the tested data files can be found in the "examples/data/" directory on [9]'s repository.

In Figure 6, we show the result that the Guest party gets which includes PSI-CA and PSU-CA. The result indicates Guest and Host has 5000 items in PSI and 5000 items in PSU.

| | dataset | intersect_count | intersect_rate | union_count |
|---|---|---|---|---|
| intersectionCA | train | 5000 | 1 | 5000 |

Figure 6. Final Output for PSI-CA and PSU-CA.

In Figure 7, we show the result that the Host party gets which is PSI: the table in this figure showed part of the intersection items.

Figure 7. Results of PSI.

The whole PSI-CA/PSU-CA and PSI protocol finished in 7.81s.

## 5. CONCLUSION

In this paper, we pointed out some mistakes of a popular PSI-CA protocol and its extended PSI protocol.

Since the original protocol's computations didn't happen in the chosen cyclic group, the protocol turned out to be wrong as well as its security proof.

After figuring out what's wrong with these protocols, we made some corrections by mapping the intermediate results to the chosen cyclic group. After the corrections, we also presented a new correctness analysis and security proof of the corrected PSI-CA protocol.

To further back up our corrected protocols, we ran a python simulation of the PSI-CA protocol and implemented PSI-CA/PSU-CA, PSI protocols in a popular open-source federated learning framework FATE. Turned out, our new corrected protocols worked perfectly well.

**Authors**

Yang Tan  Received his B.Eng. and Ph.D. from South China University of Technology. He now works as a cryptography researcher at Shenzhen Qianhai Xinxin Digital Technology Co.,Ltd and his research interests include applied cryptography, secure computation, federated learning, blockchain, etc.

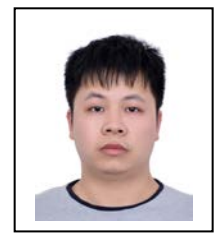

Bo Lv  Received her B.Eng. and Ph.D. from South China University of Technology. She now works as a lecturer at Huizhou University and her research interests include public key cryptography, secure multi-party computation, etc.

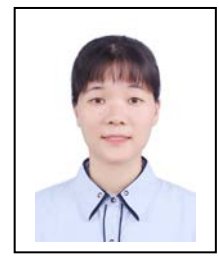